\documentclass{rmaa}
\usepackage{amsmath}

\title{The photoevaporation of a neutral structure by an
EUV+FUV radiation field}

\author{V. Lora$^1$, M. J. Vasconcelos$^2$, \\
A. C. Raga$^3$, A. Esquivel$^3$ and A. H. Cerqueira$^2$
\affil{$^1$Astronomisches Rechen-Institut, Heidelberg, Germany\\
$^2$Laborat\'orio de Astrof\'{\i}sica Te\'orica e Observacional, Bahia, Brazil\\
$^3$Instituto de Ciencias Nucleares, UNAM, M\'exico}}

\fulladdresses{
\item V. Lora: Astronomisches Rechen-Institut Zentrum f\"ur
Astronomie der Universit\"at
Heidelberg, M\"onchhofstr. 12-14, 69120 Heidelberg, Germany
(vlora@ari.uni-heidelberg.de)
\item M. J. Vasconcelos,A. H. Cerqueira : Laborat\'orio de Astrof\'{\i}sica 
Te\'orica e Observacional, DCET-UESC, Rodovia Ilh\'eus-Itabuna,
 km. 16 - Ilh\'eus, Bahia, Brazil - CEP 45662-000 (mjvasc,hoth@uesc.br)
\item A. C. Raga, A. Esquivel:
Instituto de Ciencias Nucleares, Universidad
Nacional Aut\'onoma de M\'exico, Ap. 70-543, 04510 D. F., M\'exico
(raga,esquivel@nucleares.unam.mx)}

\shortauthor{Lora et al.}
\shorttitle{EUV+FUV photoevaporation}

\SetVolume{00} \SetFirstPage{001} \SetYear{2013}
\ReceivedDate{\today} 
\AcceptedDate{Year Month Day} 

\resumen{
La radiaci\'on fotoionizante EUV y la radiaci\'on fotodisociante FUV de
estrellas nacientes, fotoevaporan la nube donde nacen, llevando a la formaci\'on
de grumos densos que eventualmente podr\'ian formar estrellas adicionales.
Estudiamos los efectos de incluir un flujo fotodisociante FUV, en modelos
de fragmentaci\'on de una nube molecular fotoevaporada y autogravitante.
Llevamos a cabo simulaciones 3D de la interacci\'on de una nube inhomog\'enea,
neutra y autogravitante, con una radiaci\'on externa de campos EUV y FUV, y
calculamos el n\'umero y la masa de los grumos que est\'an en colapso. 
Encontramos que la presencia de una regi\'on de fotodisociaci\'on entre la 
regi\'on  H II y la nube molecular tiene un efecto importante
en la fromaci\'on de estructuras densas debido a la expansi\'on de una regi\'on  
HII. En particular, incluir un campo FUV lleva a una formaci\'on mas temprana
de un n\'umero mayor de grumos densos, los cuales podr\'ian llevar a la formaci\'on
de un n\'umero mayor de estrellas.}

\abstract{
The EUV photoionizing radiation and FUV dissociating radiation 
from newly born stars photoevaporate their parental 
neutral cloud, leading to the formation of dense clumps that could
eventually form additional stars.
We study the effects of including a photodissociating FUV flux
in models of the fragmentation of a photoevaporating, self-gravitating
molecular cloud.
We compute 3D simulations of the interaction of an inhomogeneous,
neutral, self-gravitating cloud with external EUV and FUV radiation
fields, and calculate the number of collapsing clumps and their mass.
We find that the presence of an outer photodissociation region has an important 
effect on the formation of dense structures due to the expansion of
an HII region. In particular,
including a FUV field leads to the earlier formation of
a larger number of dense clumps, which might lead
to the formation of more stars.}


\begin{document}

\maketitle

\section{Introduction}
The expansion of an H~II region into a surrounding, inhomogeneous
molecular cloud leads to the formation of complex, elongated
``elephant trunk'' structures. Examples of this are the
trunks in the Eagle \citep{hester:96} and the Rosette
\citep{carlqvist:03} nebulae. Examples of similar (but larger)
structures in external galaxies have been found as well,
\citep{carlqvist:10}.

A considerable amount of theoretical work has focused on the
photoevaporation of a single, dense clump, starting with the
paper of \citeauthor{oort:55} (\citeyear{oort:55}). This problem has been addressed
both analytically \citep{bertoldi:89,bertoldi:90} 
and numerically \citep{lefloch:94,dale1:07,dale2:07,dale:11,ercolano:12}. Several simulations have
been done including detailed treatments of the radiative transfer 
and ionization \citep{mellema:98,raga:09}, and the study 
of clumps with low amplitude density inhomogeneities \citep{gonzalez:05},
with self-gravity \citep{esquivel:07} and with magnetic fields
\citep{henney:09}. These models are applicable to photoionized
regions in which individual neutral clumps are clearly visible
(the evident example of this being the Helix Nebula, see, e.~g.,
\citeauthor{odell:05} \citeyear{odell:05}).

However, the observations of elephant trunks suggest the presence
of more complex density structures in the region with neutral gas.
Attempts to address this have included 3D simulations of the propagation
of ionization fronts into ``multi-clump'' structures
\citep{lim:03,raga:09,mackey:10,mackey:11}
and into more complex density distributions \citep{mellema:06,esquivel:07,maclow:07,grithschneder:09,ercolano:11,arthur:11}.
These simulations 
took into account the following physical processes:
\begin{itemize}
\item direct gas dynamics+ionizing radiation transfer
\citep{lim:03,mackey:10},
\item self-gravity \citep{esquivel:07,maclow:07,grithschneder:10,ercolano:11},
\item magnetic fields \citep{mackey:11,arthur:11},
\item the diffuse ionizing radiation field \citep{raga:09,ercolano:11}.
\end{itemize}

An important effect that has not been explored in detail 
until very recently is the presence of a photodissociating
FUV radiation field (together with the photoionizing EUV radiation).
\citeauthor{arthur:11} (\citeyear{arthur:11}) presented numerical simulations
including the FUV radiation field and obtained that the FUV field can 
have a clear dynamical importance in the formation of dense clumps at the edge of an
expanding H~II region. Their simulations include the presence of
a magnetic field, but do not include the self-gravity of the gas.

In the present paper, we discuss 3D simulations which include the
transfer of the EUV and FUV radiation in a self-gravitating medium
with an inhomogeneous initial density distribution. Our simulations
differ from those of \citeauthor{arthur:11} (\citeyear{arthur:11}) in that they do
not include a magnetic field, but do consider the self-gravity of
the gas (see section 2).

We then use the results of our simulations to calculate the number
of clumps (section 3) and mass
distributions of the dense clumps (section 4). Finally,
the results are summarized in section 5.

\section{The numerical simulations}

We have carried out twenty four 3D simulations with the code described
by \citeauthor{lora:09} (\citeyear{lora:09}). This code integrates the gasdynamic equations
in a uniform 3D, Cartesian grid, together with the radiative transfer
of radiation at the Lyman limit, and a hydrogen ionization rate equation,
including the self-gravity of the gas.

The radiative transfer and hydrogen ionization are solved as
follows. We place an ionizing photon source (producing $S_*$
ionizing photons per unit time) far away, outside the computational
grid, and then impose an ionizing photon flux $F_0=S_*/(4\pi R_0^2)$
on the boundary of the computational grid, where $R_0$ is the distance
from the grid boundary to the photon source (assumed to lie along the
$x$-axis). The ionizing photon flux is then marched into the
computational domain as:
\begin{equation}
F_{i+1,j,k}=F_{i,j,k}\,e^{-\Delta \tau_{Ly}}\,,
\label{fp}
\end{equation}
where $F_{i,j,k}$ is the ionizing photon flux at the left boundary
(along the $x$-axis) of the $(i,j,k)$ computational cell and
\begin{equation}
\Delta \tau_{Ly}=n_{HI}\left(\sigma_{H,\nu_0}+\sigma_d\right)\Delta x\,,
\label{dtau}
\end{equation}
where $n_{HI}$ is the neutral H density of cell $(i,j,k)$, $\Delta x$
is the size of the cell (along the $x$-axis),
$\sigma_{H,\nu_0}=6.30\times 10^{-18}$~cm$^2$ is the Lyman limit
photoionization cross section of HI and $\sigma_d=1.1$~cm$^2$
is the FUV/EUV dust absorption cross section. This value of $\sigma_d$
is derived assuming that the EUV/FUV dust absorption
is $A={1.2\times 10^{-21}\rm~cm^{-2}}\,N_{HI}$ (the derivation
of this relation is discussed by \citeauthor{vasconcelos:11} \citeyear{vasconcelos:11}).
We should note that in calculating the Lyman limit optical depth through equation
(\ref{dtau}) we are assuming that the region of photoionized H has no dust. If it
did, one would have to replace $n_{HI}$ by $n_H$ (the total H number density) in
equation (\ref{dtau}). However, because of the low column densities of the ionized
H regions within our computational domain, it makes no difference whether or not
dust is present in these regions.

With the ionizing photon flux $F$, we calculate the H photoionization rate
\begin{equation}
\phi_H=\left(1-e^{-\Delta \tau_{Ly}}\right)F\,.
\label{phih}
\end{equation}
This photoionization rate is then included in a HI continuity equation:
\begin{equation}
\frac{\partial n_{HI}}{\partial t}+ \dot\nabla (n_{HI}\underline{u})=
n_en_{HII}\alpha_H(T)-n_{HI}\phi_H\,,
\label{nhi}
\end{equation}
where $n_{HI}$, $n_{HII}$ and $n_e$ are the neutral H, ionized H and electron densities
(respectivey) and $\alpha_H(T)$ is the recombination coefficient of H. This equation
is integrated together with the standard 3D gasdynamic equations. The diffuse
radiation is included only by considering the ``case B'' recombination (to all levels
with energy quantum number $N>1$).

To equations (\ref{fp}-\ref{nhi}), which were included in the code
of \citeauthor{lora:09} (\citeyear{lora:09}), we have now
added the transfer of FUV radiation, and an ionization rate equation
for CI. This has been done in the following way.
For the FUV photons, we solve the radiative transfer problem
at $\lambda=1100$~\AA\ (the ionization edge of the ground state of CI),
considering the absorption due to CI photoionization and to dust
extinction. We therefore calculate the optical depth for the FUV
photons in each computational cell as:
\begin{equation}
\Delta \tau_{FUV}=\left(n_{CI}\sigma_{C,\nu_0}+n_{HI}\sigma_d\right)\Delta x\,,
\label{tfuv}
\end{equation}
where $n_{CI}$ is the CI number density and $\sigma_{C,\nu_0}=1.22\times 10^{-17}$~cm$^2$
is the photoionization cross section at the CI ionization edge. We have
assumed that the dust extinction cross section (per H atom) has the same
values at the CI and HI ionization edges (i.e. and $\lambda\sim 1100$ and 900~\AA,
respectively). With this value for the FUV optical depth of the computational
cells, we then use equations equivalent to (\ref{fp}) and (\ref{phih}) to calculate
the CI photoionization rate. We then integrate a continuity equation for CI (with the
appropriate source terms, such as equation \ref{nhi} but for CI) together with
the gasdynamic and HI continuity equations.

Instead of integrating an energy equation (with appropriate heating and cooling
terms), we compute the temperature of the gas as:
\begin{equation}
T=(T_1-T_2)x_{HII}+T_2x_{CII}+T_3(1-x_{CII})\,,
\label{t}
\end{equation}
where $x_{HII}$ and $x_{CII}$ are the H and C ionization
fractions (respectively), and $T_1=10^4$~K, $T_2=10^3$~K
and $T_3=10$~K are the typical temperatures of photoionized,
photodissociated and molecular regions respectively.
Therefore, we do not calculate the photodissociation of H$_2$, and
assume following \citeauthor{richling:00} (\citeyear{richling:00}),
that it approximately follows the ionization of CI.

From the paper of \citeauthor{diaz:98} (\citeyear{diaz:98}), we now take the EUV
($\lambda>912$~\AA) photon rate $S_I$ and the FUV
(912~\AA$<\lambda<$1100~\AA) photon
rate $S_D$ for a set of three main sequence O stars (of effective
temperatures $T=50000$, 45000 and 40000~K). For the EUV photons,
we solve the radiative transfer problem (parallel to the $x$-axis
of the computational grid) using the Lyman-limit H absorption
coefficient $\sigma_{\nu_0}(H)=6.30\times 10^{-18}$~cm$^2$, and
using the EUV flux to calculate the HI photoionization rate
(this is completely equivalent to the models of \citeauthor{lora:09} \citeyear{lora:09}).

For our simulations, we consider a computational domain
of $(3.0,\,1.5,\,1.5)\times 10^{18}$~cm (along
the $x$-, $y$- and $z$-axes, respectively) resolved with
$256\times 128\times 128$ grid points. An outflow boundary
condition is applied on the $x$-axis boundaries, and reflection
conditions in all of the other boundaries.

This domain is initially filled with an inhomogeneous density
structure with a power law power-spectrum index of $-11/3$
(i.e., $P(k)\propto k^{-11/3}$, where $k$ is the wave number,
see \citeauthor{esquivel:03} \citeyear{esquivel:03}). This results in a density distribution
with a dispersion of $\approx 2$ times the mean density. We
have chosen four different realizations of the density
distribution, which we use to compute models 
identified with the letters M, O, D and E (each letter corresponding
to one of the chosen initial density distributions).

The medium is initially at rest.
The computational domain is divided
into an initially ionized region (with ionized H and C)
for $x<x_0=4\times 10^{17}$~cm and a neutral region
(with neutral H and C) for $x>x_0$. The average
density in the neutral medium is 100 times the average
density in the ionized medium, and the transition between
the two follows a tanh profile with a width
of $\sim 10$ pixels. The resulting neutral structure has
a total mass of 228~M$_\odot$. This initial setup is identical to
the one used by \citeauthor{lora:09} (\citeyear{lora:09}).

Other authors have used more complex initial conditions for this
kind of simulation, in particular, including an initial velocity field.
This was done, e.g. by \citeauthor{arthur:11} (\citeyear{arthur:11}), who
took as initial conditions the output from a 3D, turbulent cloud simulation.
Other examples of initial conditions can be found in \citeauthor{dale:11} (\citeyear{dale:11}), 
\citeauthor{dale:12} (\citeyear{dale:12}) and \citeauthor{ercolano:12} (\citeyear{ercolano:12}). 
However, as the flow motions induced by the photoionization and photoevaporation
have velocities which are much larger than the ones of the initial turbulent
motions, including these initial motions is unlikely to produce large
effects on the results.

We assume that we have a stellar source situated $3\times 10^{18}$~cm
from the edge of the computational domain in the $-x$ direction. For
this photon source we consider three possibilities, an O3 ($T_{eff}=50000$~K),
an O5.5 ($T_{eff}=45000$~K) and an O7.5 ($T_{eff}=40000$~K) main sequence
star. The EUV and FUV fluxes computed by \citeauthor{diaz:98} (\citeyear{diaz:98}) 
for such stars are given in Table~1.

We then compute simulations with the EUV and FUV fluxes given
in Table 1 (models M1-3, O1-3, D1-3 and E1-3, with letter M-E
corresponding to the four different initial density distributions, see
above), and simulations with the
same EUV fluxes but with zero FUV flux (models
M1B-3B, O1B-3B, D1B-3B and E1B-3B). As can be seen from Table 1, the
models with the same number (second character of the model identification)
have the same impinging EUV field.
The set of models with zero FUV fields (models M1B-E3B)
has a transition from $T_3=10$~K in the region
with neutral H to $T_1=10^4$~K in the photoionized region (see
equation \ref{t}), with no intervening photodissociated region,
and are therefore equivalent to the models of \citeauthor{lora:09} (\citeyear{lora:09}).

\section{Results}
In Figures 1 and 2, we show the mid-plane density stratifications
and the positions of the H and C ionization fronts
obtained for a $t=2.5$~kyr (Figure 1) and $t=100$~kyr (Figure 2)
integration time
for models M1, O1, D1 and E1. In Figures 3 and 4, we show the
same density stratifications, obtained for a $t=2.5$~kyr (Figure 3)
and $t=100$~kyr (Figure 4) integration time
for the models with zero FUV fluxes: M1B, O1B, D1B and E1B.

In the models with non-zero FUV fluxes (see Figures 1 and 2)
we see that the C and H ionization fronts are separated by a
photodissociated region (with CII and HI) with a width of $\sim
10^{18}$~cm in the $t=2.5$ and $100$~kyr frames. The neutral region
(to the right of the C ionization front) develops progressively
denser regions which collapse under the combined effects
of the compression ahead of the C ionization front and the
self-gravity of the gas. As expected, the higher temperature,
photodissociated region does not develop such dense structures.

For the models with zero FUV fluxes (models M1B-E1B see Figures 3 and 4)
there is of course no photodissociated
region. 

Qualitatively similar time-evolutions are obtained for all
of the computed models. Of course, the details of the fragmentation
of the neutral gas into dense clumps differ in all simulations.
These differences are quantified in the following subsection.

\begin{table}
\caption{Model Parameters}
\label{table:1}      
\centering
\begin{tabular}{lcccc}
\hline\hline\noalign{\smallskip}
Models & $T_{eff}$ & $\log_{10} S_I$ & $\log_{10} S_D$ & $S_D/S_I$\\
      & [$10^3$~K] & \multispan2{\hfil [photons~s$^{-1}$]\hfil} & \\
\hline\noalign{\smallskip}
M1, O1, D1, E1$^{\mathrm{a,b}}$ & 50 & 49.89 & 49.54 & 0.45 \\
M2, O2, D2, E2$^{\mathrm{a,b}}$ & 45 & 49.35 & 49.16 & 0.65 \\
M3, O3, D3, E3$^{\mathrm{a,b}}$ & 40 & 48.78 & 48.76 & 0.96 \\
\hline
\end{tabular}
\begin{list}{}{}
\item[$^{\mathrm{a}}$] models with letters M through E have identical
FUV and EUV fluxes, and correspond to different initial
density distributions (see the text),
\item[$^{\mathrm{b}}$] models M1B-E1B, M2B-E2B and M3B-E3B have the same values
of $S_I$ as models M1-E1, M2-E2 and M3-E3 (respectively), but have $S_D=0$.
\end{list}
\end{table}

\begin{figure}[!t]
\centering
\includegraphics[scale=0.55]{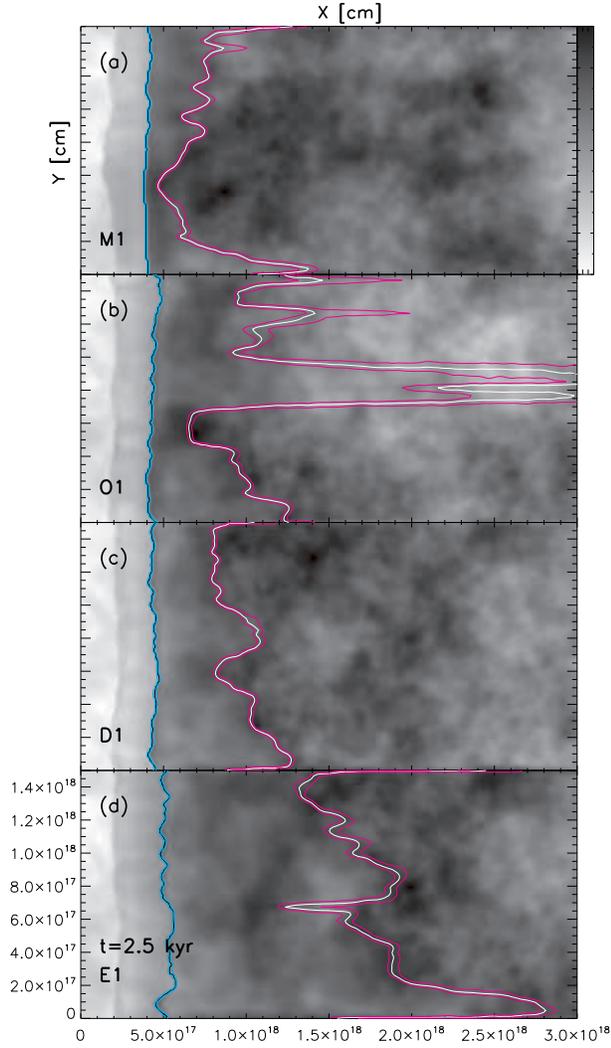}
\caption{The $t=2.5$~kyr, $xy$-mid-plane density stratifications
of models M1, O1, D1 and E1 (with non-zero FUV fields, see Table 1). 
The density stratifications are shown with the logarithmic gray scale 
given (in g~cm$^{-3}$) by the top right bar. In the four frames, we show 
the contour corresponding to an H ionization fraction of $50\%$ (black 
line), which indicates the position of the HI/II ionization front. The 
blue lines show the width of the HI/II region. The contour corresponding to 
a C ionization fraction (white line) of $50\%$, which indicates the position 
of the CI/II ionization front, is also shown. The pink lines show the width 
of the CI/II region. The $x$ and $y$-axes are labeled in cm.}
\label{fig1}
\end{figure}

\begin{figure}[!t]
\centering
\includegraphics[scale=0.55]{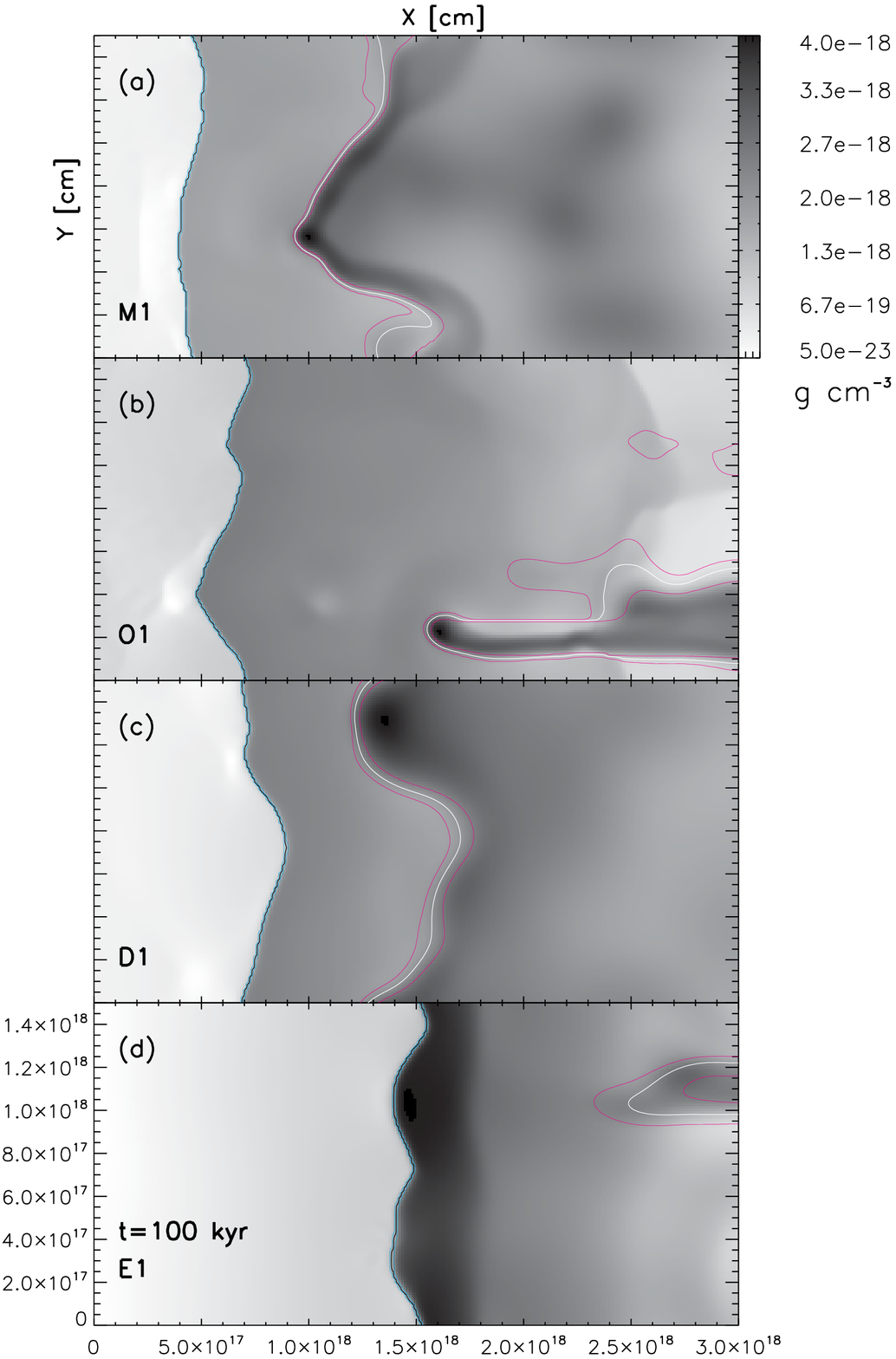}
\caption{Same as Figure 1, but for $t=100$~kyr (see Table 1).}
\end{figure}

\begin{figure}[!t]
\centering
\includegraphics[scale=0.55]{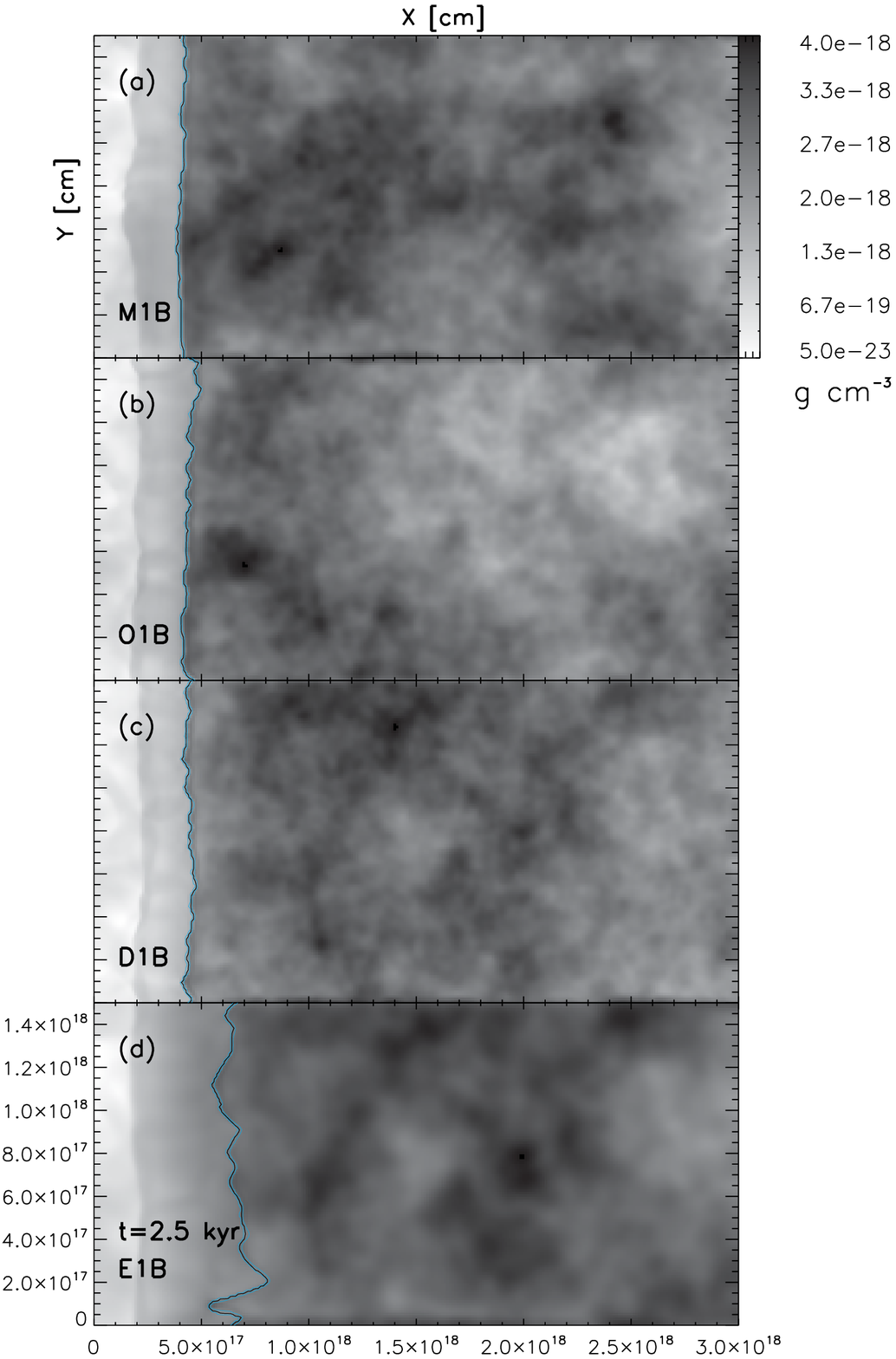}
\caption{The $t=2.5$~kyr, $xy$-mid-plane density stratifications
of models M1B, O1B, D1B and E1B (with zero FUV fields,
see Table 1). The density stratifications are shown 
with the logarithmic gray scale given (in g~cm$^{-3}$) by the top right 
bar. In the four frames, we show the contour corresponding to an H 
ionization fraction of $50\%$ (black line), which indicates the position 
of the HI/II ionization front, and also the 
contour corresponding to a C ionization fraction (white line) of $50\%$, which 
indicates the position of the CI/II ionization front. The 
blue lines show the width of the HI/II region. The $x$ and $y$-axes 
are labeled in cm.}
\end{figure}

\begin{figure}[!t]
\centering
\includegraphics[scale=0.55]{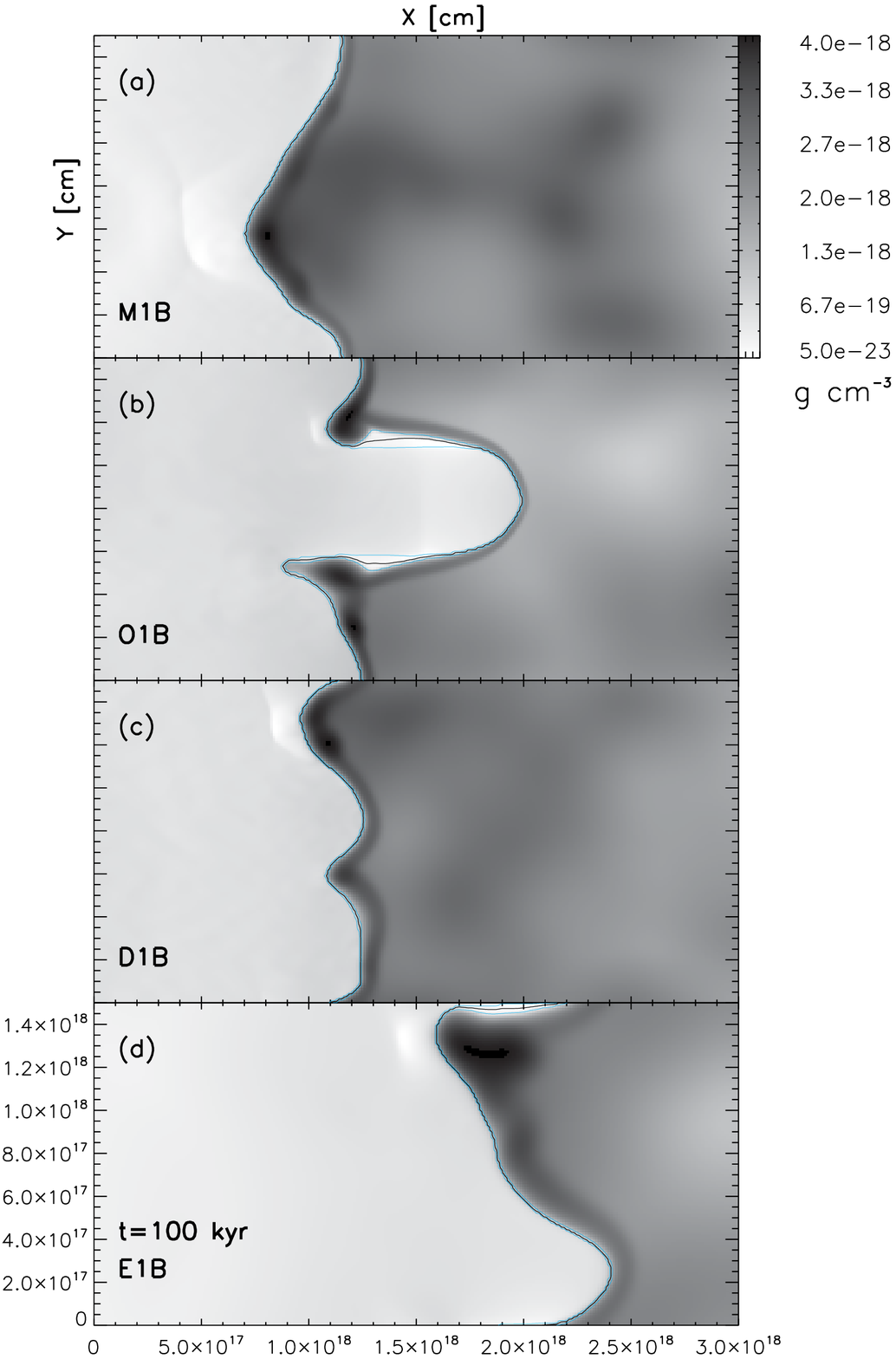}
\caption{Same as Figure 3, but for $t=100$~kyr (see Table 1).}
\end{figure}

\subsection{Number of clumps as a function of time}
We take the density stratifications resulting from
our 24 simulations (see Table 1), and compute the number
of clumps present at different integration times. To
calculate the number of clumps, we define a cutoff
density $\rho_c$, and count all spatially contiguous
structures with densities $\rho\geq \rho_c$. 

Following \citeauthor{lora:09} (\citeyear{lora:09}), 
we choose three different cutoff densities
$\rho_c=10^{-20}$, $10^{-19}$ and $3\times 10^{-18}$~g~cm$^{-3}$
(corresponding to number densities of atomic nuclei
$\sim$4600, 46000 and $1.4\times 10^6$~cm$^{-3}$). 

For the models without a FUV field (models M1B-E3B),
we count clumps in the neutral H region. For the models with
non-zero FUV fields (models M1-E3, see Table 1), we
count clumps that satisfy one of the two following conditions~:
\begin{enumerate}
\item that their material has neutral H,
\item that they have neutral C.
\end{enumerate}
Notably, these two criteria result in identical clump numbers
for the $\rho_c=10^{-19}$ and $3\times 10^{-18}$~g~cm$^{-3}$
cutoff densities (see Figures 5, 6 and 7). For $\rho_c=10^{-20}$~g~cm$^{-3}$, the
second criterion results in somewhat larger clump numbers.
In the computation of the clump numbers, we calculate
averages over each of the sets of four models with identical
parameters (see Table 1) but with different initial
density structures.

\begin{figure}[!t]
   \centering
   \includegraphics[scale=0.5]{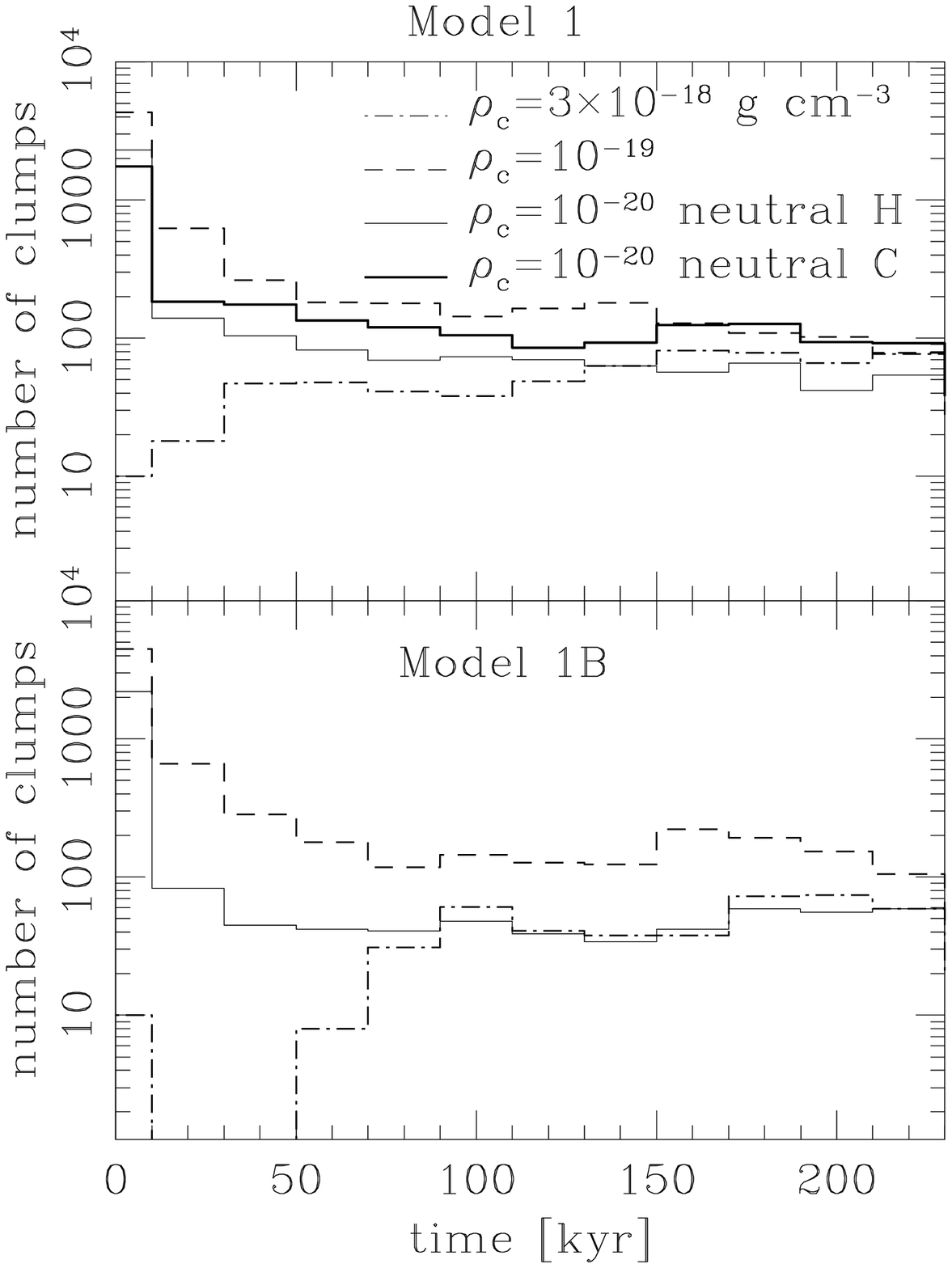}
\caption{The top panel shows the number of neutral 
clumps obtained from models M1, O1, D1 and E1
(with non-zero FUV fluxes, see Table 1) as a function of time, 
for the three chosen density cutoffs. The bottom
panel shows the number of neutral clumps for the same density cutoffs but for
models M1B, O1B, D1B and E1B (with zero FUV, see Table 1).
In the top panel, the two solid lines correspond
to a $\rho_c=10^{-20}$~g~cm$^{-3}$ cutoff density, with clumps with neutral
H (thin line) and with neutral C (thick line).
}
\end{figure}

\begin{figure}[!t]
   \centering
   \includegraphics[scale=0.5]{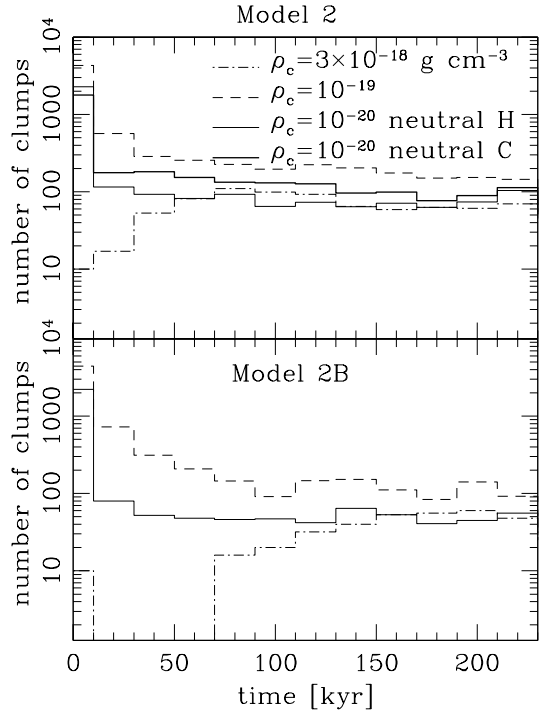}
\caption{Same as Figure 5, but for models M2, O2, D2 and E2 (top frame)
and M2B, O2B, D2B and E2B (bottom frame).}
\end{figure}

\begin{figure}[!t]
   \centering
   \includegraphics[scale=0.5]{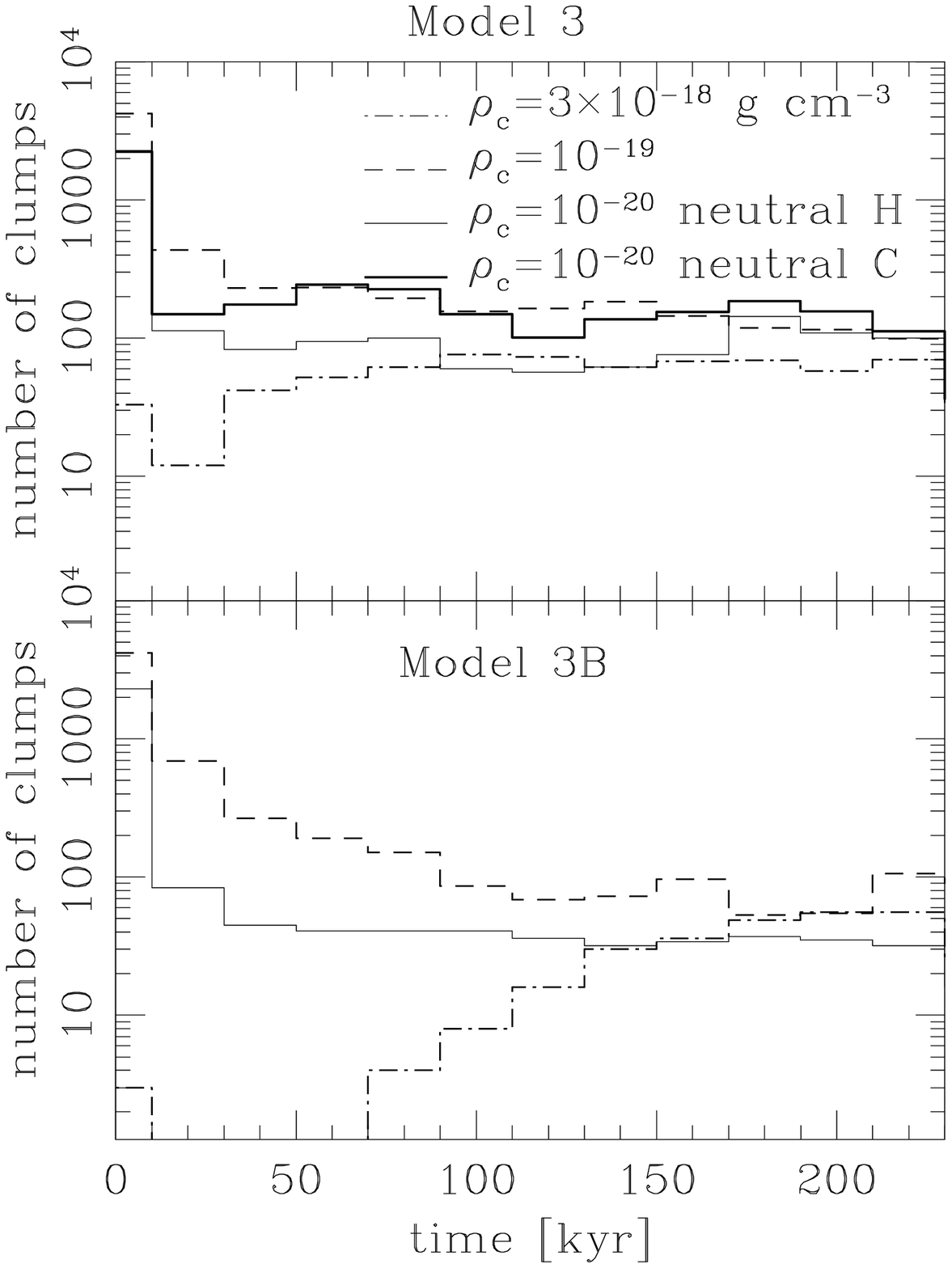}
\caption{Same as Figure 5, but for models M3, O3, D3 and E3 (top frame)
and M3B, O3B, D3B and E3B (bottom frame).}
\end{figure}

The results obtained from this clump counting exercise are
given in Figures 5 (models M1-E1 and M1B-E1B), 6 (models M2-E2
and M2B-E2B) and 7 (models M3-E3 and M3B-E3B). These Figures show the number of
clumps (averaged over 20 kyr time-intervals) as a function
of the integration time, obtained with the three chosen cutoff densities.

From Figures 5-7, we see that for the $\rho_c=10^{-20}$~g~cm$^{-3}$ cutoff
density, the number of clumps first decreases rapidly,
and then stabilizes (for $t>20$~kyr) at a value of $\approx 100$
for most of the models. Actually, if for the models with
nonzero FUV field we count clumps with neutral C,
the number of clumps stabilizes at a value of $\sim 200$ (see
the top panels of Figures 5-7). 

For the $\rho_c=10^{-19}$~g~cm$^{-3}$ cutoff
density, the number of clumps starts at a value of $\sim 3\times10^3$,
and in all models decreases to $\sim 200$ within the first
$\sim 50$~kyr of the time-evolution. For larger times, in
the models with zero FUV flux (M1B-E1B, M2B-E2B and M3B-E3B, bottom panels
of Figures 5, 6 and 7, respectively), the clump number decreases
and then stabilizes (for $t>30$~kyr) at a value of $\approx 50$.

For $\rho_c=3\times 10^{-18}$~g~cm$^{-3}$, the models
with zero FUV flux take $\sim 50\to 70$~kyr to develop the first clump,
and by $t=130$~kyr they have developed $\sim 40$ clumps. The models with
non-zero FUV flux (top panels of Figures 5-7) develop the first
clump much earlier, after only $\sim 10$~kyr, and have about
$70$ clumps at $t=130$~kyr.

Even though our models with zero FUV flux (M1B, M2B and M3B)
cover a factor of $\sim 13$
in EUV photon rates (see Table 1), except for relatively small effects
(e.g., the earlier appearance of high $\rho_c$ clumps in model M1B-E1B,
see the bottom panel of Figure 5) they show qualitatively similar
trends of number of clumps as a function of time. Our models with
non-zero FUV flux (M1-E1, M2-E2 and M3-E3, see Table 1) also cover a factor
of $\sim 6$ of FUV photon production rates, and these three
models also show
qualitatively similar time evolution of the number of clumps
(top panels of Figures 5-7).

From this, we conclude that for the range of early to late O-type stars
chosen, the fragmentation of the neutral structure into
clumps presents a similar behavior regardless of the spectral subclass of the star.
We also see that including the effect of the FUV photons does produce
important differences (see the above list of 3 items). The effect
of introducing the FUV flux is 
to increase the number of $\rho_c = 3\times 10^{-18}$~g~cm$^{-3}$ clumps
(from $\sim 40$ to $\sim 70$) at the final, $t=240$~kyr integration
time of our simulations.

\begin{figure}[!t]
   \centering
   \includegraphics[scale=0.4]{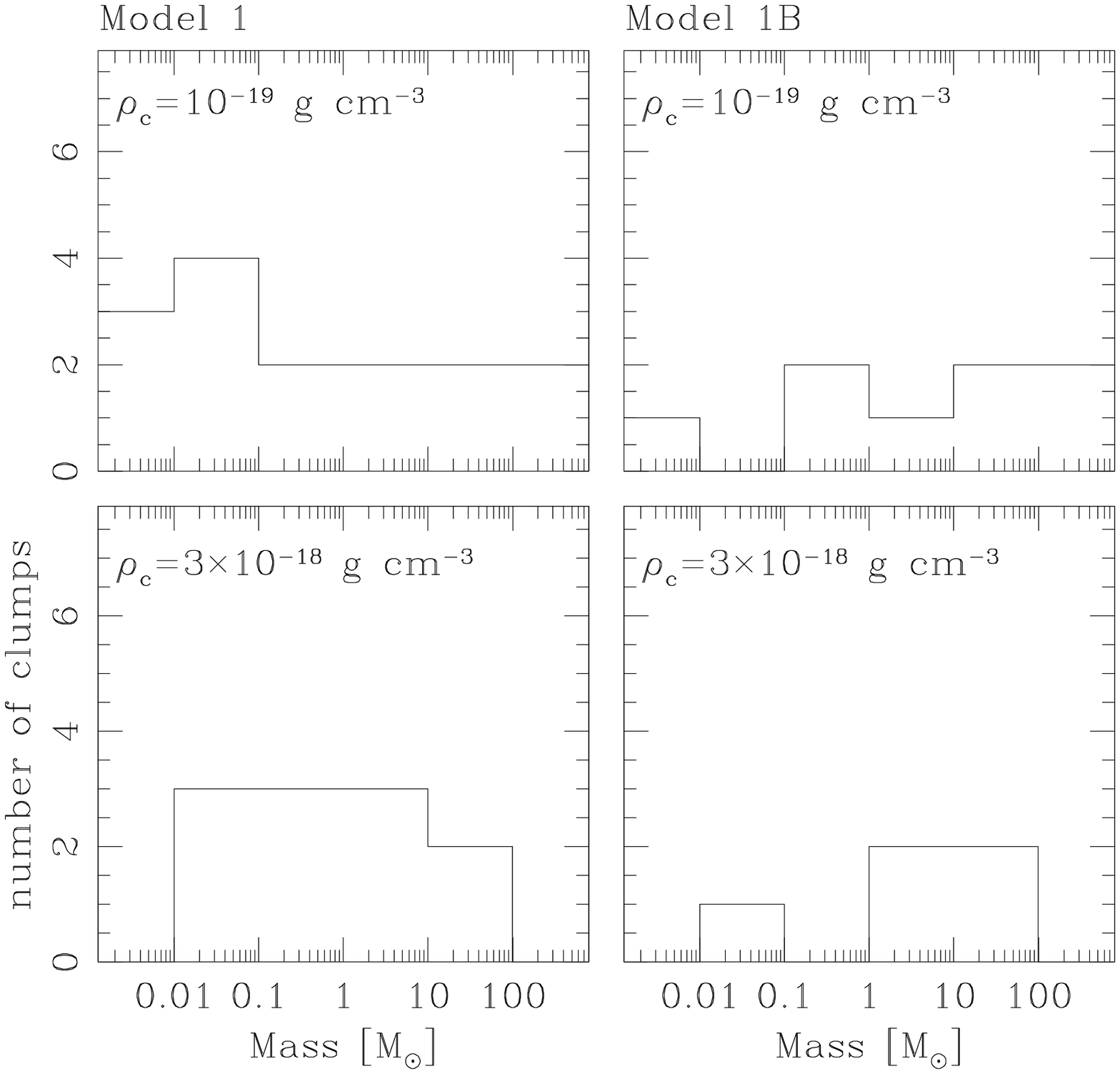}
\caption{Neutral clump mass distribution for two cutoff densities
($\rho_c=10^{-19}$ and $3\times10^{-18}$~g~cm$^{-3}$) for the integration
time $t=150$~kyr for models M1, O1, D1 and E1 (left panels)
and M1B, O1B, D1B and E1B (right panels). }
\end{figure}

\begin{figure}[!t]
   \centering
   \includegraphics[scale=0.4]{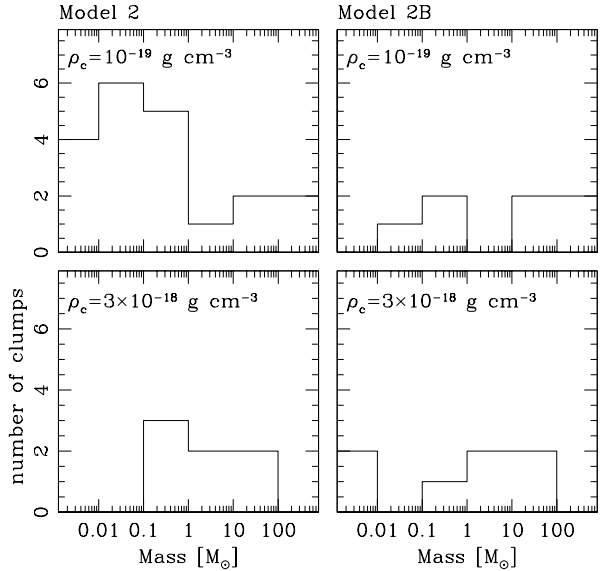}
\caption{Same as Figure 8, but for models M2, O2, D2 and E2 (left
panels) and M2B, O2B, D2B and E2B (right panels).}
\end{figure}

\begin{figure}[!t]
   \centering
   \includegraphics[scale=0.4]{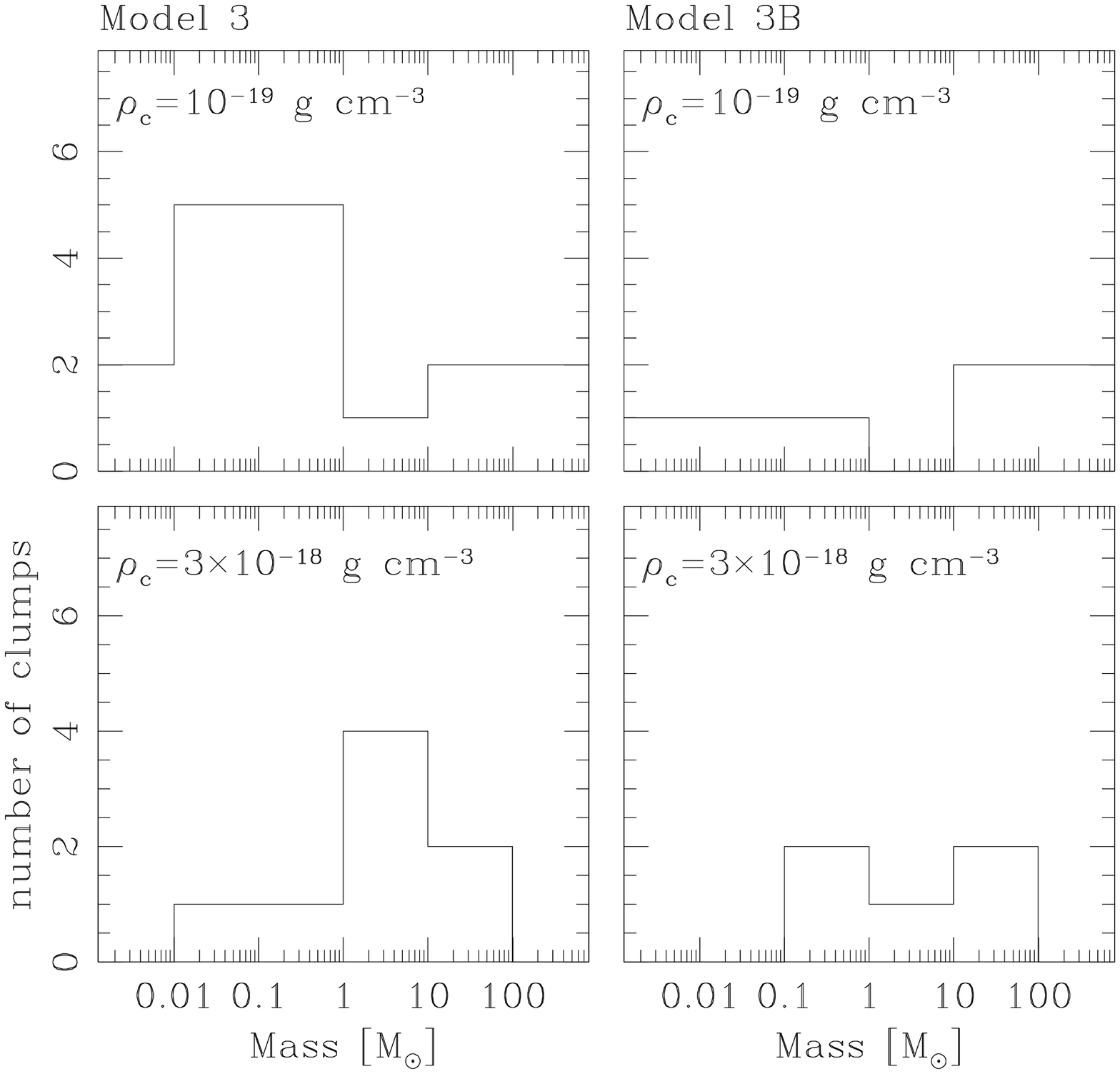}
\caption{Same as Figure 8, but for models M3, O3, D3 and E3 (left
panels) and M3B, O3B, D3B and E3B (right panels).}
\end{figure}

\subsection{The clump mass distributions}
We now focus on a $t=150$~kyr evolutionary time, in which
a sizable population of clumps has developed in all models
(see Figures 5-7). For the stratifications resulting from
all models at this time, we compute the mass distributions
of the clumps obtained with the $\rho_c=10^{-19}$ and
$3\times 10^{-18}$~g~cm$^{-3}$ cutoff densities. We do
not compute the distributions for $\rho_c=10^{-20}$~g~cm$^{-3}$
because at $t=150$~kyr they only have one clump which includes
basically all of the neutral region of the flow.

The resulting clump number vs. mass distributions are shown
in Figures 8 (models M1-E1 and M1B-E1B, see Table 1), 9 (models M2-E2 and M2B-E2B)
and 10 (models M3-E3 and M3B-E3B). If we look at the $\rho_c=10^{-19}$~g~cm$^{-3}$
clump mass distributions (top panels of Figures 8, 9 and 10), we observe that the 
distributions of all models have two clumps in the
high mass, $10^2\to 10^3$~M$_\odot$ bin, and for models with non zero FUV flux 2-4 clumps
in the low mass, $10^{-3}\to 10^{-2}$~M$_\odot$ bin.

The distributions of the models with zero FUV flux
(M1B-E1B, M2B-E2B and M3B-E3B) have 0-1 clump in the $10^{-2}\to 10^{-1}$~M$_\odot$
range, while the distributions of the non-zero FUV flux
models (M1-E1, M2-E2 and M3-E3) have 4-6 clumps in this mass
range (except for model M1, with one clump in this mass range,
see Figure 8).

Therefore, the main effect of including a non-zero FUV field
is mostly to enhance the $\rho_c=10^{-19}$~g~cm$^{-3}$ number of clumps in
the $10^{-3}\to 10^{0}$~M$_\odot$ mass range.

If we look at the $\rho_c=3\times 10^{-18}$~g~cm$^{-3}$
clump mass distributions (bottom panels of Figures 8, 9 and 10), we see that
the three models with zero FUV flux (M1B-E1B, M2B-E2B and M3B-E3B) show
very similar clump mass distributions, with 
one clump in the $10^{-1}\to 10^2$~M$_\odot$ mass range for
model M1B-E1B, and no clumps in this mass range for models
M2B-E2B and M3B-E3B.
In the $10^{-2}\to 10^{-1}$~M$_\odot$
range, all of the models with non-zero FUV flux have 7-8 clumps,
generally having 3 more
clumps in this range than the zero FUV models (4-5 clumps).

Therefore, the presence of an FUV flux allows the formation
of more low mass, $\rho_c=3\times 10^{-18}$~g~cm$^{-3}$ clumps,
than in the zero FUV flux models.

\section{Summary}
In a previous paper \citep{lora:09}, we have studied
the formation of dense clumps in the interaction of a
photoionizing radiation field with an inhomogeneous medium
(with an initial power law spectrum of density fluctuations).
The applicability of these models to real astrophysical flows
(associated with expanding HII regions) was questionable because
of the absence (in the models) of a photodissociation region
preceding the HI/II ionization front.

In this work, we present a set of numerical simulations
which explore the effect of a FUV radiative field, which
produces a photodissociation region outside the HII region.
To this effect, we compute 3D simulations which include
the photoionization of H and C, assuming that the CI/II ionization
front approximately coincides with the outer edge of the
photodissociation region (as initially suggested by
\citeauthor{richling:00} \citeyear{richling:00}). \citeauthor{arthur:11} (\citeyear{arthur:11}) calculate
numerical simulations in which a different approximation
is used for determining the outer edge of the photodissociation
region, but which leads to similar results.

From our simulations, we obtain the masses and the number of
clumps (defined as contiguous regions of density higher than
a given cutoff density $\rho_c$, see section 3). We calculate
models with photoionizing/photodissociating stars at $\approx 1$~pc
from the computed region, and explore a range from
early to late-type O stars (using the EUV and FUV photon
rates presented by \citeauthor{diaz:98} \citeyear{diaz:98}). We also
compute models setting the FUV flux to zero, in order
to isolate the effects of having a non-zero FUV flux.

We find that including the photodissociation
region produced by a non-zero FUV flux has the following
main effects:
\begin{itemize}
\item clumps with low cutoff densities ($\rho_c=10^{-19}$~g~cm$^{-3}$)
are slightly depleted (see sections 3 and 4),
\item denser clumps (with $\rho_c=3\times 10^{-18}$~g~cm$^{-3}$)
develop earlier than in the models with zero FUV (see Figures
5-7),
\item a larger number of dense clumps
(with $\rho>\rho_c=3\times 10^{-18}$~g~cm$^{-3}$)
is produced, and these clumps have a broader mass distribution than
in the zero FUV models (see section 4).
\end{itemize}

From this, we conclude that the presence of an outer
photodissociation region has an important effect on the
formation of dense structures due to the expansion of
an HII region in an initially inhomogeneous medium. In particular,
including a FUV field leads to the earlier formation of
a larger number of dense clumps. This in principle, might lead
to the formation of more young stars. However, our simulations
do not have the resolution nor include the physical processes
necessary for determining whether or not the clumps actually
collapse to form one or more stars.

\acknowledgments
VL gratefully acknowledges support from the Alexander von
Humboldt Foundation fellowship and H.B-L.
We acknowledge support from the CONACyT grant 61547, 101356 and
101975.
AHC would like to thank the CNPq for partial financial
support (307036/2009-0).
The authors would like to also thank the anonymous referee, for very useful comments.

\end{document}